\documentclass[aps,twocolumn,pra,showpacs,floatfix]{revtex4}
\usepackage{epsfig}
\usepackage{dcolumn}

\begin{document}

\title{\bf Limits on gravitational Einstein Equivalence  Principle violation from monitoring atomic clock 
frequencies during a year} 

\author{V. A. Dzuba$^1$ and V. V. Flambaum$^{1,2}$}
\affiliation{$^1$School of Physics, University of New South Wales,
Sydney 2052, Australia}
\affiliation{$^2$Helmholtz Institute Mainz, Johannes Gutenberg University, , 55099 Mainz, Germany}

\date{\today}

\begin{abstract}
Sun's gravitation potential at earth varies during a year due to varying Earth-Sun distance. Comparing
the results of very accurate measurements of atomic clock transitions performed at different time in
the year allows us to study the dependence of the atomic frequencies on the gravitational potential.
We examine the measurement data  for the ratio of the frequencies in Hg$^+$ and Al$^+$
clock transitions and absolute frequency measurements (with respect to caesium frequency standard)
for Dy, Sr, H,  hyperfine transitions in Rb and H, and obtain significantly improved limits on the values of the gravity related parameter of the 
Einstein Equivalence  Principle violating term in the Standard Model Extension Hamiltonian $c_{00} = (3.0 \pm 5.7) \times 10^{-7}$ and 
the  parameter  for the gravity-related
variation of the fine structure constant $\kappa_{\alpha} = (-5.3 \pm 10) \times 10^{-8}$.
\end{abstract} 
\pacs{06.20.Jr, 06.30.Ft, 31.15.A, 32.30.Jc }
\maketitle


Theories unifying gravity with other interactions suggest that local Lorentz invariance (LLI) and Einstein
equivalence principle (EEP) might not be exact at very high energy~\cite{Alan89}.
This can manifest itself at low energy via tiny change of atomic frequencies,
variation of fundamental constants, etc. One can search for the LLI or EEP violation in electron sector taking 
advantage of extremely high accuracy of atomic clocks or using a system where the effect is strongly enhanced. 
The strongest constrain on the gravity related EEP violation has been obtained with the use atomic 
dysprosium~\cite{LLI-Dy13}. The Standard Model Extension (SME) 
Hamiltonian~\cite{Alan97} is used to quantify the results of the measurements. 
Corresponding term  can be presented 
in the form (see, e.g.~\cite{Saf-Nature})
\begin{equation}
\delta H = c_{00}\frac{2}{3}\frac{U}{c^2} \frac{p^2}{2m},
\label{e:dH}
\end{equation}
where $ c_{00}$ is  the  parameter characterising the magnitude of EEP violation, $U$ is the gravitation potential,
$c$ is the speed of light, $p$ is the operator of the electron momentum ($p=-i\hbar \nabla$). 
The change of the frequency of atomic transition between states $a$ and $b$ between two dates in the year is
\begin{eqnarray}
&&\Delta \omega_{ab} = c_{00}\frac{2}{3}\frac{\Delta U}{c^2} \left[ \left \langle \frac{p^2}{2m} \right \rangle_a-
\left \langle \frac{p^2}{2m} \right \rangle_b \right]  \nonumber \\
&&\equiv c_{00}\frac{2}{3}\frac{\Delta U}{c^2}\delta K_{ab}.
\label{e:domega}
\end{eqnarray}
To avoid any confusion with the sign let us assume that state $a$ is always above state $b$ on the energy scale so that
$\hbar \omega_{ab} =E_a-E_b>0$.
 $\Delta U$ in (\ref{e:domega}) is the change of the Sun's gravitation potential due to changing of the Earth-Sun distance,
$\langle \frac{p^2}{2m} \rangle_a$ is the expectation value of the kinetic energy of electrons in state $a$, and
$\delta K_{ab}$ is the difference between the  kinetic energies of the states $a$ and $b$.
The maximal change of the gravitation potential is between January and July,
$\Delta U/c^2 \approx 3.3\times 10^{-10}$~\cite{Flam07,Dy-LLI}. Therefore, comparing accurate frequency 
measurements performed in January and July, or fitting several measurements with a cosine function 
with the zero phase  in the beginning of January and with period of one year, one can put constrains  on 
the parameter $c_{00}$,
\begin{equation}
c_{00} = \frac{3}{2}\frac{\Delta \omega}{(\Delta U/c^2)\delta K_{ab}}.
\label{e:c00}
\end{equation}
Measuring atomic frequency means comparing it to some reference frequency, e.g. caesium primary frequency
standard or another microwave or optical reference frequency. Therefore, we need to consider a ratio
of two frequencies.
In the non-relativistic limit one can use the Virial theorem
($\langle p^2/2m\rangle = - E_{\rm total}$) and  
 obtain from Eq. (\ref{e:domega})
\begin{equation}
\frac{\Delta \omega}{\omega} = -c_{00}\frac{2}{3}\frac{\Delta U}{c^2},
\label{e:nr}
\end{equation}
i.e. $\Delta \omega/\omega$ is the same for all electron transitions 
(except for the hyperfine transitions where the splitting is due to  the "relativistic" magnetic interaction). 
This means that in the non-relativistic limit the effect in the ratio of optical frequencies is unobservable.
Therefore, we should  perform relativistic calculations.
It is convenient to introduce  relativistic factors $R$ which describe deviation of the expectation value of the  kinetic energy
from the value, given by the Virial theorem,
\begin{equation}
R=-\frac{\Delta E_a -\Delta E_b}{E_a-E_b}.
\label{e:R}
\end{equation}
Here $\Delta E_a$ is the energy shift of the state $a$ due to the kinetic energy operator. In the non-relativistic limit
$R=1$. In the relativistic case $R$ can be larger or smaller than one and can even be negative.
For the relative change of two frequencies we now have
\begin{equation}
\frac{\Delta \omega_1}{\omega_1} - \frac{\Delta \omega_2}{\omega_2} = 
(R_2-R_1)\frac{2}{3}c_{00}\frac{\Delta U}{c^2} \equiv (\beta_1 - \beta_2)\frac{\Delta U}{c^2}.
\label{e:R1R2}
\end{equation}
 It is clear from (\ref{e:R1R2}) that for higher sensitivity
one should compare the frequencies of atomic transitions with the largest possible difference in the
values of relativistic factors $R$.
It is convenient to rewrite (\ref{e:R1R2}) in a form
\begin{equation}
c_{00} = \frac{3}{2}\frac{\Delta \omega_1/\omega_1 - \Delta \omega_2/\omega_2} 
{(R_2-R_1)\Delta U/c^2}.
\label{e:c00R1R2}
\end{equation}
It is useful to note that in the case of $R_1 \gg R_2$ (e.g., Dy vs Cs, see below)  we come back to Eq.~(\ref{e:c00}).

If the change of atomic frequencies is fitted to a cosine function $A\cos(2\pi t/1{\rm yr})$ then one can 
obtain from (\ref{e:R1R2})
\begin{equation}
c_{00} = \frac{3A}{(R_2-R_1)(\Delta U/c^2)}.
\label{e:Ac00}
\end{equation}

The change of an atomic frequency can  also be attributed to the variation of the fine structure constant
$\alpha$ ($\alpha = e^2/\hbar c$)
\begin{equation}
\frac{\Delta \omega}{\omega} = K_{\alpha}\frac{\Delta \alpha}{\alpha} \equiv \frac{2q}{\omega}
\frac{\Delta \alpha}{\alpha}.
\label{e:q}
\end{equation}
Here $K_{\alpha}$ and $q$ are the electron structure factors ($K_{\alpha} = 2q/\omega$) which come
from atomic calculations. Assuming that $\alpha$ can very with the gravitation potential, one can write
\cite{Flam07}
\begin{equation}
\frac{\Delta \alpha}{\alpha} = \kappa_{\alpha}\frac{\Delta U}{c^2},
\label{e:kappa}
\end{equation}
where $\kappa_{\alpha}$ is an unknown parameter. 
Using (\ref{e:q}) and  (\ref{e:kappa}) we find
\begin{equation}
\kappa_{\alpha} = \frac{(\Delta \omega/\omega)}{K_{\alpha}(\Delta U/c^2)}.
\label{e:kappa_a}
\end{equation}
For the case of relative change of two frequencies one can write
\begin{equation}
\kappa_{\alpha} = \frac{\Delta \omega_1/\omega_1-\Delta \omega_2/\omega_2}{(K_{\alpha1}-K_{\alpha2})(\Delta U/c^2)}.
\label{e:kappa_aR}
\end{equation}

Comparing (\ref{e:c00}) and (\ref{e:kappa_a}) we see that the same experimental data on the change 
of atomic frequencies between January and July can be used to put constrains on the gravity-related 
parameter $c_{00}$ of the EEP violating Hamiltonian, and variation of the fine structure constant due to the change 
of the gravitational potential ($\kappa_{\alpha}$). In principle, the parameters $c_{00}$ and $\kappa_{\alpha}$  may have different physical origin and have different dependence on the reference frame (see e.g. Refs. \cite{Alan89,Alan97,Flam07}). However, for a convenience of the interpretation of different laboratory experiments we
can relate  two parameters to each other using Eqs. (\ref{e:c00}), (\ref{e:q}), and (\ref{e:kappa}):
\begin{equation}
c_{00} = \frac{3 q \kappa_{\alpha}}{\delta K_{ab}}.
\label{e:c00-kappa}
\end{equation}
Or for the case of two frequencies using (\ref{e:R1R2}) and (\ref{e:kappa_aR}) we get
\begin{equation}
\kappa_{\alpha} = \frac{R_2-R_1}{K_{\alpha1}-K_{\alpha2}}\frac{2}{3}c_{00}.
\label{e:kappa-c00}
\end{equation}

The study of the variation of $\alpha$ due to the change of the gravitation potential was a subject of previous 
works~\cite{Flam07,Dy-a-08,Dy-LLI,Blatt08}. In this paper we mostly focus on the EEP violating term (\ref{e:dH}).


It was shown in Ref. ~\cite{Saf-Nature} that the values of the matrix elements of the kinetic energy 
operator are very sensitive to the many-body effects.
Therefore, it is convenient to reduce the calculations to the calculation of the energies where we have accurate methods for the relativistic  many-body calculations. 
If the EEP violating operator is taken in the relativistic form, 
$2 E_K = c\gamma_0 \gamma^jp_j$, its inclusion into the calculation 
can be reduced to the simple rescaling of the kinetic energy term in the Dirac equation,
\begin{eqnarray}
&&\left(\frac{\partial f}{\partial r} + \frac{\kappa}{r}f\right)(1+s)
    -\left[ 2+\alpha^2(\epsilon - \hat V)\right]g = 0, \nonumber \\
&&\left(\frac{\partial g}{\partial r} - \frac{\kappa}{r}g\right)(1+s)
    +(\epsilon - \hat V)f = 0. \label{eq:Dir}
\end{eqnarray}
Here $s$ is the rescaling parameter, $s=0$ corresponds to the Dirac
equation with no extra operator. The value of $s$ should be chosen to
ensure linear dependence of the energy shift on $s$.
 In practice, it can be taken between $10^{-4}$
and $10^{-3}$. Potential $\hat V$ in (\ref{eq:Dir}) includes the nuclear
and electronic parts. The electronic part is usually the self-consistent
Hartree-Fock-Dirac potential. Note that  the EEP violating perturbation operator $\delta H$ 
(the parameter $s$) is included  into the self-consistent procedure and produces a correction 
to the Hartree-Fock potential (in the linear approximation in $\delta H$  this  is equivalent to the
random-phase approximation (RPA) with exchange). 

The rest of the calculations is the same
as for the energies. One can use the many-body perturbation theory
(MBPT), configuration interaction (CI) or any other technique
suitable for a many-electron atom and perform calculations for several
values of the rescaling parameter $s$, including $s=0$. Then, the
energy shift linear with respect to $s$ is extracted. 
The advantage of this approach comes from the fact that the
accuracy for the energy shift is expected to be similar to the
accuracy for the energy which may be controlled by the comparison of the calculated and experimental energies.
This is important since a strong sensitivity of
the EEP violating energy shift to the many-body effects
makes it difficult to estimate the accuracy of the calculations.

Table \ref{t:P2} shows the results of the calculations for some popular optical clock transitions.
The largest sensitivity can be achieved when monitoring the ratio of the frequencies of E2 and E3
transitions in Yb$^+$ ($\Delta R = 3.38$, see Table~\ref{t:P2}). This ratio has been recently measured 
to about $10^{-16}$ accuracy~\cite{Yb+14}. However, separate sets of data dated between January and July
(or between July and December) are not available. For other clock transitions, typical value of measured
frequency ratios is also $\sim 10^{-16}$ (see, e.g. \cite{Hg-Sr,Katori2016}), and a typical value of 
$\Delta R$ is $\Delta R \sim 0.2 - 0.8$ (see Table~\ref{t:P2}).
\begin{table}
\caption{\label{t:P2}
Relativistic factors for best optical clock transitions in atoms and ions.}
\begin{ruledtabular}
\begin{tabular}{l llcll cd}
\multicolumn{1}{c}{Atom/Ion}& 
\multicolumn{5}{c}{Transition}&
\multicolumn{1}{c}{$\hbar\omega$ [cm$^{-1}$]} &
\multicolumn{1}{c}{$R$} \\
\hline
Al$^+$  & $3s^2$ & $^1$S$_0$      &-& $3s3p$ & $^3$P$^o_0$ &   37393 & 1.00 \\
Sr$^+$ &  $5s$   & $^2$S$_{1/2}$ &-& $4d$ & $^2$D$_{5/2}$   &   14556  & 1.20 \\
Sr         & $5s^2$ & $^1$S$_0$      &-& $5s5p$ & $^3$P$^o_0$ &   14317 & 1.03 \\
Yb        & $6s^2$ & $^1$S$_0$      &-& $6s6p$ & $^3$P$^o_0$  &  17288  & 1.20 \\
Yb$^+$ & $6s$   & $^2$S$_{1/2}$  &-& $5d$ & $^2$D$_{3/2}$    &   22961  & 1.48 \\
Yb$^+$ & $6s$   & $^2$S$_{1/2}$  &-& $4f$  & $^2$F$_{7/2}$     &   21419  & -1.9 \\
Hg         & $6s^2$ & $^1$S$_0$     &-& $6s6p$ & $^3$P$^o_0$ &  37645  & 1.40 \\
Hg$^+$  & $5d^{10}6s$ & $^2$S$_{1/2}$  &-& $5d^9 6s^2$ & $^2$D$_{5/2}$  &   35515  & 0.2 \\
\end{tabular}
\end{ruledtabular}
\end{table}

Several extra steps are needed to calculate the relativistic factors for microwave transitions, e.g. the hyperfine
transition in caesium which serves as the primary frequency standard. First, the wave functions are found 
using Eq. (\ref{eq:Dir}) with the rescaled kinetic energy operator. Second, some standard technique is used 
to calculate the hyperfine structure including the many-body corrections.
We use the correlation potential method (see, e.g.~\cite{DzuFlaSus84,DzuFlaSus85,DzuFlaSilSus87})
to calculate the relativistic factors for the hyperfine structure of the ground state of all alkali atoms from 
Li to Cs. This ab initio method provides accuracy $\sim 1\%$ for the hyperfine structure of the alkali atoms.
The method includes solving the Hartree-Fock-Dirac equations in an external hyperfine filed (equivalent to the random phase 
approximation (RPA)) to account for the core polarization effect. The RPA equations are similar to Eq.  (\ref{eq:Dir})
but with the right-hand side containing the hyperfine interaction operator. The kinetic energy terms in the RPA 
equations  have also  been rescaled. As for the energies, the calculations are done 
for several values of the rescaling parameter $s$.
The calculated relativistic factors are presented in Table~\ref{t:hfs}. 
In the non-relativistic limit the effect of the kinetic energy rescaling results in $R=2.5$ for the hyperfine splitting (the kinetic energy scales as $1/(1+s)^2$, the hyperfine structure as $1/(1+s)^5$).

\begin{table}
\caption{\label{t:hfs}
Relativistic factors for the hyperfine clock transitions in atoms.}
\begin{ruledtabular}
\begin{tabular}{ccccccc}
& H & Li & Na & K & Rb & Cs \\
\hline
$Z$ &  1 & 3      &   11 &    19 &  37 & 55 \\
$R$ & 2.50 & 2.50 & 2.51 & 2.53 &  2.66 & 2.89 \\
\end{tabular}
\end{ruledtabular}
\end{table}


\paragraph{Al$^+$ vs Hg$^+$.}

\begin{figure}
\epsfig{figure=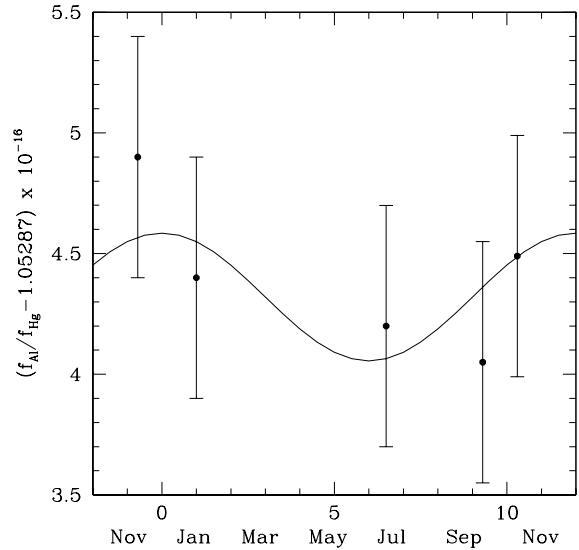,scale=0.4}
\caption{Fit by $A\cos(\omega t)+B$ the results of the measurements of the frequency ratio for the 
clock transitions in Al$^+$ and Hg$^+$ \cite{Rosenband2008} between November 2006 and November 2007.}
\label{f:Hgit}
\end{figure}

Fig. \ref{f:Hgit} shows the results of the measurements of the ratio of frequencies of Al$^+$ and Hg$^+$
clock transitions performed between December 2006 and November 2007 in Ref. ~\cite{Rosenband2008}.
Note that we approximately reproduce Fig. 3A of this paper. The measurements are fitted by the
cosine function $A\cos(2\pi t/1{\rm yr})+B$. The least mean square fitting 
leads to $A=(0.26 \pm 0.50) \times 10^{-16}$, $B=(4.3 \pm 0.4)\times 10^{-16}$. Substituting $A$ into Eq. 
(\ref{e:Ac00}) and using the relativistic factors $R_1=1$  for Al$^+$ and $R_2=0.2$  for Hg$^+$ (see Table \ref{t:P2})
we obtain
\begin{equation}
c_{00} = (-3.0 \pm 5.7) \times 10^{-7}.
\label{e:AlHg}
\end{equation}
We can also use the data to extract the limit on the gravity-related variation of the fine structure constant
$\kappa_{\alpha}$. Using (\ref{e:kappa-c00}) and the values $K_{\alpha1} \approx 0$ for Al$^+$ and 
$K_{\alpha2} \approx -3$ for Hg$^+$~\cite{FlaDzuCJP} we get
\begin{equation}
\kappa_{\alpha} = (5.3 \pm 10) \times 10^{-8}.
\label{e:AlHg}
\end{equation}
This represents  5 times improvement of the previous result with dysprosium~\cite{Dy13} (see below).

\paragraph{Dysprosium.}

Dysprosium atom has a unique pair of degenerate states of the opposite parity, state A $4f^{10}6s5d$, $J=10$
and state B $4f^{9}5d^26s$, $J=10$, both having energy $E_{A,B}$=19797.96 cm$^{-1}$  above the ground state.
Due to an extremely small energy interval between these states, many effects relevant to new physics are 
strongly enhanced \cite{DzuFlaKhr86,DzuFlaWeb99,DzuFlaWeb99a}. The transition between these states was used to study parity 
non-conservation~\cite{Dy-PNC}, time variation of the fine structure constant~\cite{Dy-a-08,Dy13,LLI-Dy13}, 
LLI and EEP violation~\cite{Dy-LLI,LLI-Dy13}, and search for dark matter~\cite{Dy-DM}. The latest study of the coupling of the
variation of the fine structure constant to gravity reveals~\cite{Dy13}
\begin{equation}
\kappa_{\alpha} = (-5.5 \pm 5.2) \times 10^{-7}.
\label{e:Dyka}
\end{equation}
We can use this result together with Eq.~(\ref{e:c00-kappa}) to obtain the value of $c_{00}$. To do so we also 
need to know $\Delta q$ and $\delta K_{ab}$. Both values come from the atomic calculations. The sensitivity
coefficients $q$ for states $A$ and $B$ of Dy ($\Delta q = q_A - q_B$) were calculated in 
Refs.~\cite{DzuFlaMar03,DzuFla08a}. The values are $q_A= $ 7952 cm$^{-1}$, $q_B$= -25216 cm$^{-1}$. 
Both values are stable and reliable.  The value of $\delta K_{ab}$ was calculated in Ref.~\cite{LLI-Dy13}.
However, we believe that this number is inaccurate. Our new calculations performed as described above
 produce quite different results, $\langle p^2/2m \rangle_A$ = 57.632 a.u., and 
 $\langle p^2/2m \rangle_B$ = 57.803 a.u., leading to $\delta K_{ab}$ = 0.174 a.u. These values agree well
 with what is expected from the Virial theorem. The CI energy of ten external electrons of Dy is 
 $E^{\rm CI}_{A,B} = -61.89$ a.u. for both states A and B. Small difference between $|E^{\rm CI}|$ and
 $\langle p^2/2m \rangle$ can be attributed to the relativistic effects. In contrast, the values calculated and
 used in Ref.~\cite{LLI-Dy13} are too large and strongly disagree with the Virial theorem. The most likely
 reason for the disagreement is a bug in the computer code. In the end, our present value for $\delta K_{ab}$
 is 43 times smaller (with the minus sign) than in Ref.~\cite{LLI-Dy13}. 
 
 Substituting these numbers into Eq.~(\ref{e:c00-kappa}) we obtain
 \begin{equation}
 c_{00} = (6.4 \pm 6.0) \times 10^{-7}.
 \label{e:c00Dy}
 \end{equation}
 This result agrees with the result of Ref.~\cite{LLI-Dy13} if the latter is corrected by the factor of 43.

\paragraph{Hg$^+$ and Sr optical clocks, Rb and hydrogen.}

\begin{table}
\caption{\label{t:c00}
Limits on the EEP violating parameter $c_{00}$ from different experiments.}
\begin{ruledtabular}
\begin{tabular}{lcr}
\multicolumn{1}{c}{Clocks} & Ref. & \multicolumn{1}{c}{$c_{00}$}  \\
\hline
 Al$^+$/Hg$^+$ & \cite{Rosenband2008} &  $(-3.0 \pm 5.7) \times 10^{-7}$ \\
Dy/Cs                 & \cite{LLI-Dy13}             & $(6.4 \pm 6.0) \times 10^{-7}$ \\
Hg$^+$/Cs         & \cite{Hg+07}                 & $(1.2 \pm 2.1)\times 10^{-6}$ \\
Sr/Cs                  & \cite{Blatt08}                & $(-0.93 \pm 1.3)\times 10^{-5}$ \\
H/Cs\footnotemark[1] & \cite{H-maser}     & $(2.9 \pm 5.5)\times 10^{-5}$ \\
Rb/Cs                         & \cite{RbCs}          & $(0.7 \pm 6.8)\times 10^{-6}$ \\
H/Cs\footnotemark[2] & \cite{HCs}     & $(0.4 \pm 5.4)\times 10^{-6}$ \\
\end{tabular}
\footnotetext[1]{ $1s$ - $2s$ optical transition in H.} 
\footnotetext[2]{ Hydrogen maser.} 
\end{ruledtabular}
\end{table}

Frequency of the electric quadrupole clock transition in Hg$^+$ at $\hbar \omega = 35515$ cm$^{-1}$ 
(last line of Table~\ref{t:P2}) was measured against caesium clock to a very high precision in \cite{Hg+07}. 
The limit on the modulated change of the frequency during a year was found to be 
\begin{equation}
\frac{\Delta (\omega_{{\rm Hg}^+}/ \omega_{\rm Cs})}{ (\omega_{{\rm Hg}^+}/ \omega_{\rm Cs})}= (0.7 \pm 1.2) \times 10^{-15}.
\label{e:Hg-dw}
\end{equation}
Substituting the numbers into Eq. (\ref{e:c00R1R2}) and using the values $R=0.2$ from Table~\ref{t:P2} 
and $R=2.89$ from Table~\ref{t:hfs} we get
\begin{equation}
c_{00}=(1.2 \pm 2.1)\times 10^{-6}.
\label{e:c00-Hg}
\end{equation}
Similarly, frequency of the $^1$S$_0$ -  $^3$P$^o_0$ clock transition in Sr
was measured in \cite{Blatt08} against caesium clock  limiting the modulated change of the frequency to
\begin{equation}
\frac{\Delta (\omega_{\rm Sr}/ \omega_{\rm Cs})}{(\omega_{\rm Sr}/ \omega_{\rm Cs})}= (-3.8 \pm 5.4) \times 10^{-15}.
\label{e:Sr-dw}
\end{equation}
which leads to
\begin{equation}
c_{00}=(-0.93 \pm 1.3)\times 10^{-5}.
\label{e:c00-Sr}
\end{equation}
Two sets of measurements of the $1s$ - $2s$ energy interval in hydrogen separated by the time interval
of 44 months  puts limit on the variation of the transition 
frequency~\cite{H-maser}
\begin{equation}
\frac{\Delta (\omega_{\rm H}/ \omega_{\rm Cs})}{(\omega_{\rm H}/ \omega_{\rm Cs})} = (-12 \pm 23) \times 10^{-15} .
\label{e:H-dw}
\end{equation}
The measurements were done in June-July of 1999 (near the gravitational potential minimum) and February 
of 2003 (near the gravitational potential maximum). Then using (\ref{e:c00R1R2}) with $R=1$ for hydrogen and $R=2.89$ for caesium we get
\begin{equation}
c_{00}=(2.9 \pm 5.5)\times 10^{-5}.
\label{e:c00-H}
\end{equation}
 In Refs. \cite{RbCs,HCs} the values of parameters $\beta$ (see Eq.~(\ref{e:R1R2})) have been measured using 
 comparisons of the frequencies of the hyperfine transitions Rb/Cs and H/Cs: $\beta(Rb)-\beta(Cs)= (0.11 \pm 1.04) 
 \times 10^{-6}$ and $\beta(H)-\beta(Cs)= (0.1 \pm 1.4) \times 10^{-6}$. Using (\ref{e:R1R2}) and (\ref{e:c00R1R2})
 and the values of $R$  from Table~\ref{t:hfs} we obtain 
\begin{equation}
c_{00}=(0.7 \pm 6.8)\times 10^{-6}\,\,\,, c_{00}=(0.4 \pm 5.4)\times 10^{-6}
\label{e:c00-Rb}
\end{equation}
correspondingly. The results are summarized in Table~\ref{t:c00}.

We see that the best limits on the value of $c_{00}$ come from the frequency measurements in Al$^+$ vs
Hg$^+$ (\ref{e:AlHg}) and Dy vs Cs (\ref{e:c00Dy}), they  stand at the $10^{-7}$ level.  The current fractional
accuracy for the  Al$^+$ vs Hg$^+$ measurements is $10^{-16}$~\cite{Rosenband2008}. Since optical
clocks approach fractional accuracy of $10^{-18}$ (see, e.g. ~\cite{clock-18a,clock-18b,clock-18c,clock-18d}) 
further progress in limiting the gravity-dependent EEP violating interaction is possible. Many existing 
measurements can probably be used for this purpose if the date of the measurements is known.

Note that corresponding measurements are also sensitive to the EEP violation in the electromagnetic  interaction. 
We will consider this effect in a separate publication.

The work was supported in part by the Australian Research Council and the Gutenberg Fellowship at Mainz University.
The authors would like to  thank  D. Budker for useful comments and  Mainz Institute for Theoretical Physics (MITP)
for its hospitality and support.

\bibliographystyle{apsrev}


\end{document}